# Exponential Trust Based Mechanism to Detect Black Hole attack in Wireless Sensor Network

Dr. Deepali Virmani [1], Manas Hemrajani [2], Shringarica Chandel [3]

*Bhagwan Parshuram Institute of Technology,*

deepalivirmani@gmail.com[1]  manas.hemrajani@gmail.com[2]  shringaricachandel37@gmail.com[3]

*Abstract*-Security is a key feature in Wireless Sensor Networks but they are prone to many kinds of attacks and one of them is Black Hole Attack. In a black hole attack all the packets are consecutively dropped which leads to the decrease in the efficiency of the network and unnecessary wastage of battery life. In this paper, we propose an exponential trust based mechanism to detect the malicious node. In the proposed method a Streak counter is deployed to store the consecutive number of packets dropped and a trust factor is maintained for each node. The trust factor drops exponentially with each consecutive packet dropped which helps in detecting the malicious node. The proposed method show a drastic decrease in the number of packets dropped before the node being detected as a malicious node.

## I. INTRODUCTION

Wireless Sensor Networks consists of a number of sensors which monitor physical environment condition and collaboratively pass data to the destination. [1] In recent years, security of wireless sensor networks has been a major issue. Wireless sensor network are vulnerable to many types of threats. One of the major and the most serious threat is a black hole attack. A black hole attack is carried out by the malicious nodes in the network which exploits the routing protocol to advertise itself as having the shortest path to the node whose data it wants to obstruct[2]. The black hole attack poses severe threat to the security of wireless sensor network since it drops packets continuously. In this paper we attempt to find out the malicious node in the network using exponential trust based mechanism. We use the concept of overhearing which was proposed in the watchdog mechanism [5]. Cluster heads are formed to perform the overhearing task. A streak counter is maintained in the system to keep count of the consecutive number of packets dropped by a particular node. Trust factor is also maintained for each node in the table which falls exponentially when a node drops packets. When the trust factor goes below a threshold trust value then the node is declared as malicious and as the one carrying out black hole attack in the network.

### A. Black Hole Attack

Black hole is a severe attack against the routing protocol of the sensor network. In this attack the malicious node counterfeits other nodes by announcing a shortest false route to the destination. The malicious node then attracts additional traffic and drops the packets continuously. [3] Black hole attack is a type of denial of service attack. When the source node wishes to transmit data to the destination it sends a Route REQuest (RREQ) message to all the nodes. Malicious nosed also being a part of the network receive RREQ message and replies with Route REPly (RREP) message ahead of all the other nodes. It thus attracts additional traffic to it falsely claiming the shortest route to the destination and drops them continuously.

Remainder of the paper is organized as follows. Section 2 briefly states related works for detection of malicious nodes in the network. Section 3 describes our proposed model. In section 4 we present our proposed algorithm and flowchart. In section 5 we give the advantages of our proposed algorithm and finally conclude our paper in section 6.

## II. RELATED WORK

There are number of mechanism proposed for detection of malicious nodes in the network. Some of them are described below briefly.

In[4], Yuanming Wu et. al. discussed security vulnerabilities of watchdog mechanism[5] and trust mechanism and also examined how inside attackers could exploit them. The work was based on detection of inside attackers and their trust mechanism involved three stages: 1) node behavior monitoring, 2) trust measurement, and 3) insider attack detection.

In [5], Marti Guili et. al. proposed a watchdog mechanism technique which worked on the concept of eavesdropping. The node can overhear all the transmissions within its radial transmission range.

The watchdog mechanism had many drawbacks due to its simple overhearing method. This mechanism was improved by A. Forootaniniaand [6] where a cluster head node was fixed and buffers were used to store packets sent by the nodes.

In [7], S.Banerjee et. al. has proposed an algorithm to overcome black hole and gray hole attacks. The algorithm proposes to divide the complete data into smaller blocks and rely on the network to find out and eliminate the malicious nodes. The adjacent nodes cooperate in the delivery of the small blocks from the source to the destination. The acknowledgement from the destination helps in the detection of malicious node.

In [8], Jian yin et. al. proposed a hierarchal secure routing protocol against black hole attack which used symmetric key cryptography to discover a secure route against black hole attacks. In hierarchal secure routing protocol network is divided into different groups organized as a tree with each group leader as the root of the tree. After inter-group shared key and intra-group shared key establishment, black hole attacks are detected locally. For detection of cooperative black hole attack a randomized data acknowledgement scheme is proposed.

### III. PROPOSED MODEL

In the Exponential Trust based mechanism, we maintain a table in the memory which stores the trust factor (*TF*) of each node. Initially, trust factor is 100 for every node. A streak counter is maintained in memory which keeps count of every consecutive packet dropped. Initially the value for each node of the counter is 0 and is incremented by 1 with every successive packet dropped. As soon as the node forwards a packet to its next node in the routing table the streak counter is set to 0 again. Here, we use the fact that the node carrying out black hole attack is going to drop all the packets. The trust factor of a node is calculated by the formula $100(x^{n_i})$. The exponential drop in the *TF* with the increase in the streak counter is shown in Table I.

The fault tolerance of the network is also considered. If fault tolerance is very high for the network then it should be kept closer to 1 when the network can tolerate packets being dropped. So the decrease in the TF with each consecutive packet dropped will be very less. If it is closer to 0 that means the fault tolerance of the network is very less. In this case, the trust factor of the node will fall drastically for every node. Value of x should hence, be chosen carefully. If the value of fault tolerance is too high it can lead to a large number of packets being dropped before being detected and a very low value will lead to the node being declared as malicious after dropping only a few consecutive packets.

The fall in TF is rapid as the streak counter is in the exponent as can be seen in the fig 1. The fault tolerance taken for the graph is 0.95.

Table. I.

Values of trust factor varying with the consecutive number of packets

| Consecutive Packets dropped | x | Trust factor 100(x^n) |
|---|---|---|
| 1 | 0.95 | 95 |
| 10 | 0.95 | 59.87369392 |
| 20 | 0.95 | 35.84859224 |
| 30 | 0.95 | 21.46387639 |
| 40 | 0.95 | 12.85121566 |
| 50 | 0.95 | 7.694497528 |
| 60 | 0.95 | 4.606979899 |
| 70 | 0.95 | 2.758369044 |
| 80 | 0.95 | 1.651537439 |
| 90 | 0.95 | 0.988836471 |
| 100 | 0.95 | 0.592052922 |
| 200 | 0.95 | 0.003505267 |
| 300 | 0.95 | 2.0753E-05 |
| 500 | 0.95 | 7.27449E-10 |
| 1000 | 0.95 | 5.29182E-21 |

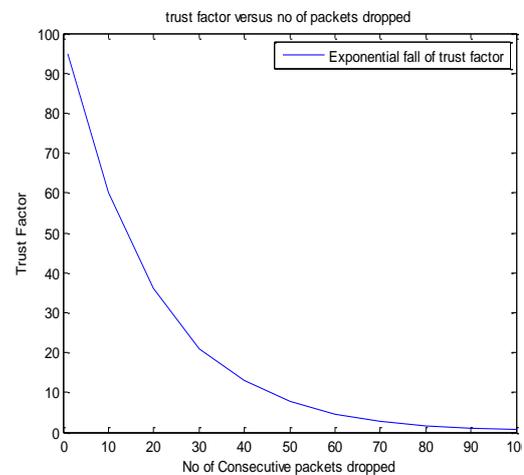

Fig. 1. Trust factor versus number of packets dropped

### IV. PROPOSED ALGORITHM

1. S is taken as the Source node.
2. $T_e$ is the Energy Timer which is initialised when a new node is selected as the Cluster Head Node.
3. $T_r$ is a timer which is initialized when a node receives a packet. The timer is set to

a real time delay according to the tolerance of the network.
4. M is the Cluster Head Node which is selected on the basis of Energy. All the nodes broadcast their energy after $T_e$ goes off. The node with the maximum available energy is selected as the cluster head.
5. ACK is the Acknowledgement that will be sent by the Head Node when the packet reaches its destination.
6. RTR is the request to resend packet sent by the cluster head node when a packet fails to reach the destination.
7. $n_i$ is the consecutive number of packets dropped by $i^{th}$ node.
8. $TF_i$ is the Trust Factor of $i^{th}$ node. It is calculated as $100(x^{\wedge}n_i)$. As $n_i$ increases, the $TF_i$ decreases exponentially as shown in table.
9. x is the fault tolerance of the network.
10. TTF is the Threshold Trust Factor. Any $TF_i$ going below the TTF results in $i^{th}$ node being declared as malicious.

---

1. All nodes broadcast their energy E.
2. $i^{th}$ node with Max Energy ($E_i$) is selected as Head node (M) in a cluster to overhear all the transmissions.
3. S forwards packet to its neighbour in routing table.
4. M overhears the packet being sent to Node.
5. INT T=real time delay
6. if the next node = destination then
   6.1 Send ACK to the Source node.
7. else if next node = Intermediate node
   7.1 If node forwards the packet within real time delay
   7.2 Set $n_i$=0.
8. else if node doesn't forward the packet within real time delay
   8.1 Set $n_i$=$n_i$+1
   8.2 Compute Trust Factor of $i^{th}$ node. $TF_i$ = $100(x^{\wedge}n_i)$ where $n_i$ is the Streak counter
   8.3 if $TF_i$ <= TTF (threshold trust factor) value then broadcast $i^{th}$ node as malicious (carrying out black hole attack).
   8.4 Send RTR to the source node
   8.5 Go to step 1

---

The flowchart is shown on the next page in fig. 2.

## V. ADVANTAGES OF PROPOSED ALGORITHM

1. Our model doesn't depend on total number of packets dropped by a node. It takes into consideration the order in which the packets are dropped. If total number of packets dropped is considered then a node can initially forward a few packets and build its trust and then suddenly drop a large number of packets. It can avoid being detected by staying above the threshold value.
2. It prevents to an extent the cooperative black hole attack which is carried out by two or more
3. The model considers a fault tolerance factor which enables it to be deployed in various topologies and applications.
4. If the node is not malicious in a network and becomes malicious after some time then also the model can detect the attack being carried.
5. The dynamic clustering takes into account that the overheads incurred due to overhearing computations are done by the node which has maximum energy available. This makes sure energy of the nodes is used up efficiently.
6. The cluster head makes sure that the packets are retransmitted by the source node in case an intermediate node drops them.

## VI. CONCLUSION

Wireless Sensor Network have a wide range of practical applications. But they are vulnerable to many types of attacks. One of the Attacks is a Black Hole attack. In this paper we tried to overcome Black hole attack by using Exponential Trust Based mechanism. The mechanism uses a streak counter which keeps a count of consecutive packets dropped. The mechanism is better than other methods as the trust factor falls exponentially with every successive packet dropped and have many advantages as stated in the above section.

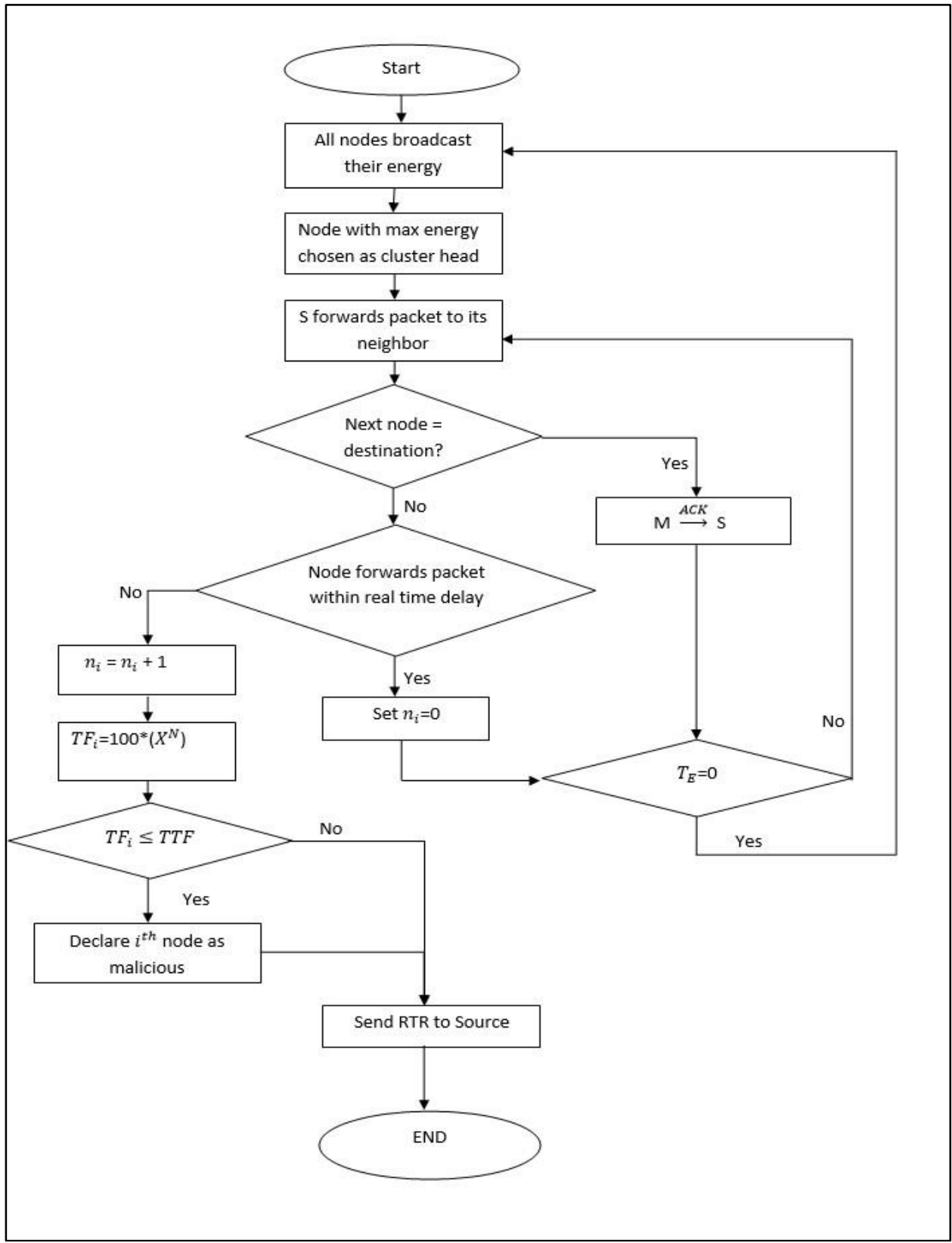

Fig 2. Flowchart